\documentclass[12pt]{iopart}
\usepackage[utf8]{inputenc}
\usepackage{graphicx}
\usepackage{subcaption}
\usepackage{hyperref}
\usepackage{xcolor}
\usepackage{verbatim}
\usepackage{booktabs}

\begin{document}
\title{Encapsulation  Structure and  Dynamics in Hypergraphs}
\author{Timothy LaRock}
\address{Mathematical Institute, University of Oxford, UK}
\ead{larock@maths.ox.ac.uk}

\author{Renaud Lambiotte}
\address{Mathematical Institure, University of Oxford, UK}
\address{Turing Institute, London, UK}
\ead{renaud.lambiotte@maths.ox.ac.uk}

\vspace{10pt}
\begin{indented}
\item[]June 2023
\end{indented}

\begin{abstract}
Hypergraphs have  emerged as a powerful modeling framework to represent systems with multiway interactions, that is systems where interactions  may involve an arbitrary number of agents. Here we explore the  properties of real-world hypergraphs, focusing on the encapsulation of their hyperedges, which is the extent that smaller hyperedges are subsets of larger hyperedges. Building on the concept of line graphs, our measures quantify the relations existing between hyperedges of different sizes and, as a byproduct, the compatibility of the data with a simplicial complex representation -- whose encapsulation would be maximum. We then turn to the impact of the observed structural patterns on diffusive dynamics, focusing on a variant of threshold models, called encapsulation dynamics, and demonstrate that non-random patterns can accelerate the spreading in the system.
\end{abstract}

\vspace{2pc}

\section{Introduction}

Networks provide a powerful language to model and analyze interconnected systems \cite{newman2018networks}. The building blocks of networks are pairwise edges, and these blocks can then be combined to form walks and paths, making it possible for systems to be globally connected yet sparse. Since the seminal work of Watts and Strogatz 25 years ago \cite{watts1998collective}, a key focus of network science has been to investigate the relationship between the structure of a network and the dynamics taking place on its nodes \cite{lambiotte2021modularity}. This program requires the design of metrics to capture significant, non-random structural properties of  networks, {\it e.g.}, the clustering coefficient, the degree distribution or modularity, as well as the specification of dynamical models, both linear and non-linear, for the diffusion between neighbouring nodes. An important observation is that the same structural property may affect different dynamical models in different ways, {\it e.g.}, a high density of triangles tends to slow down simple diffusion, but facilitate complex diffusion \cite{guilbeault2018complex}. 

Finding the right modeling framework for interacting systems is a challenging task. While networks have the advantage of simplicity, it has been recognized that they may also neglect critical aspects of a system and even lead to a misleading representation. Driven by the availability of datasets with richer connectivity information in recent years, different frameworks have emerged to enrich the network representation, leading to different types of higher-order networks \cite{lambiotte2019networks}. One branch of this research has extended pairwise graph-based models to multiway interaction frameworks, most notably as hypergraphs or simplicial complexes, to account for group interactions among arbitrary numbers of nodes \cite{battiston2021physics,bick2021higher,salnikov2018simplicial}.

Multiway interactions naturally appear in many systems, ranging from social interactions, where people interact in groups rather than in pairs \cite{patania2017shape}, to joint neuronal activity in brains \cite{petri2014homological} and cellular networks \cite{klamt2009hypergraphs}. Different computational tools have been adapted to multiway systems, for instance for centrality measures \cite{benson2019three} and community detection \cite{yin2017local}. Researchers have also investigated how the structure of multiway interactions impacts dynamical processes \cite{lanchier_stochastic_2013, bodo_sis_2016, iacopini2019simplicial,schaub2020random,bick2022multi}, especially the conditions under which dynamics on hypergraphs and simplicial complexes differ from those on networks \cite{neuhauser2020multibody,cencetti2023distinguishing,neuhauser_learning_2023}.

The objectives of this paper are twofold: to propose metrics that characterise the non-random patterns of {\em encapsulation} in multiway systems, and to explore dynamical models that may be affected positively or negatively by this type of hypergraph structure. These objectives are motivated by a well-known  conceptual difference between the two main representations for multiway systems, hypergraphs and simplicial complexes. By definition, a simplicial complex of size $k$ nodes includes all of the subfaces of the complex. In contrast, a hyperedge of size $k$ nodes does not imply the existence of any subsets as hyperedges in the same  hypergraph. We refer to this difference as the {\em simplex assumption}. For example, using a simplicial complex to represent the relationship between 3 nodes $\{a,b,c\}$ assumes that the subfaces $\{a,b\}, \{a,c\}, \{b,c\}$ all exist, along with the individual nodes. This is a strong assumption that is unlikely to hold, even approximately, in real data. A classic example is co-authorship, where a jointly authored paper between three co-authors does not imply that each pair of co-authors have also authored separate papers together, nor that each co-author has published a single-author paper. Recent work has investigated the relationship between these two representations \cite{baccini_weighted_2022}, and shown that the choice of higher-order representation does effect the outcome of dynamical processes \cite{petri_simplicial_2018, sun_higher-order_2021, zhang_higher-order_2023}.  

Simplicial complexes and hypergraphs can be seen as poles on a spectrum of multiway interaction structure, and it is likely that real data falls somewhere in-between. In this work, we build on previous investigations of this spectrum of overlapping higher-order structures, as well as random models for hypergraphs and simplicial complexes \cite{courtney_generalized_2016, courtney_weighted_2017, chodrow_configuration_2020, do_structural_2020, cencetti_temporal_2021-1, sun_higher-order_2021, lee_how_2021, kim_contagion_2023}. 
Our approach builds on the notion of line graph, that has been used in different contexts in network science, where nodes are the edges of the original graph and there is a link between two nodes if their corresponding edges have a node in common \cite{evans2009line, aksoy2020hypernetwork}. The interactions between hyperedges of arbitrary sizes make it possible to define a variety of different line graphs for hypergraphs. As each hyperedge can be seen as a set of nodes, this problem is equivalent to that of comparing two sets. There exist multiple ways to compare sets, which leads to multiple ways to build a line graph for a hypergraph \cite{costa2021further}.

We will focus in particular on what we refer to as an \emph{encapsulation graph}, where two hyperedges are connected (by a directed edge from larger to smaller) if one is the subset of the other. 
We then analyze the properties of the resulting directed acyclic graphs built from real-world hypergraphs  and from a synthetic hypergraph model called the Random Nested Hypergraph Model (RNHM) \cite{kim_contagion_2023}, which allows for some control over the extent of  nested structure through random rewiring of simplicial complexes. Finally, we define a process for the spread of a complex contagion on a hypergraph through its hyperedges, and show how varying levels of encapsulation structure impact the spread of the contagion in both synthetic and real hypergraphs.

\section{Measuring Overlap and Encapsulation in Hypergraphs}
Consider a list of multiway interactions, where each item in the list is a set of nodes that represent a group interaction. We will represent these interactions as a hypergraph, and focus in particular on \emph{aggregated, static} hypergraphs, where all interactions are included regardless of any dynamic or temporal information. In fact, for all of the empirical datasets we will examine, this static hypergraph is actually the result of aggregating interactions that happen over time. We will also make our hypergraphs \emph{simple}, meaning that no edges are repeated, {\it i.e.}, hyperedges are contained in a set, rather than a multiset. In the future, the techniques we develop here could be extended to study the relationships between hyperedges over time, extending, for example, work on simplicial closure \cite{benson_simplicial_2018} or temporal dynamics of group interactions \cite{kim_weighted_2017, iacopini_temporal_2023}.

\begin{figure}
    \centering
    \includegraphics[scale=0.38]{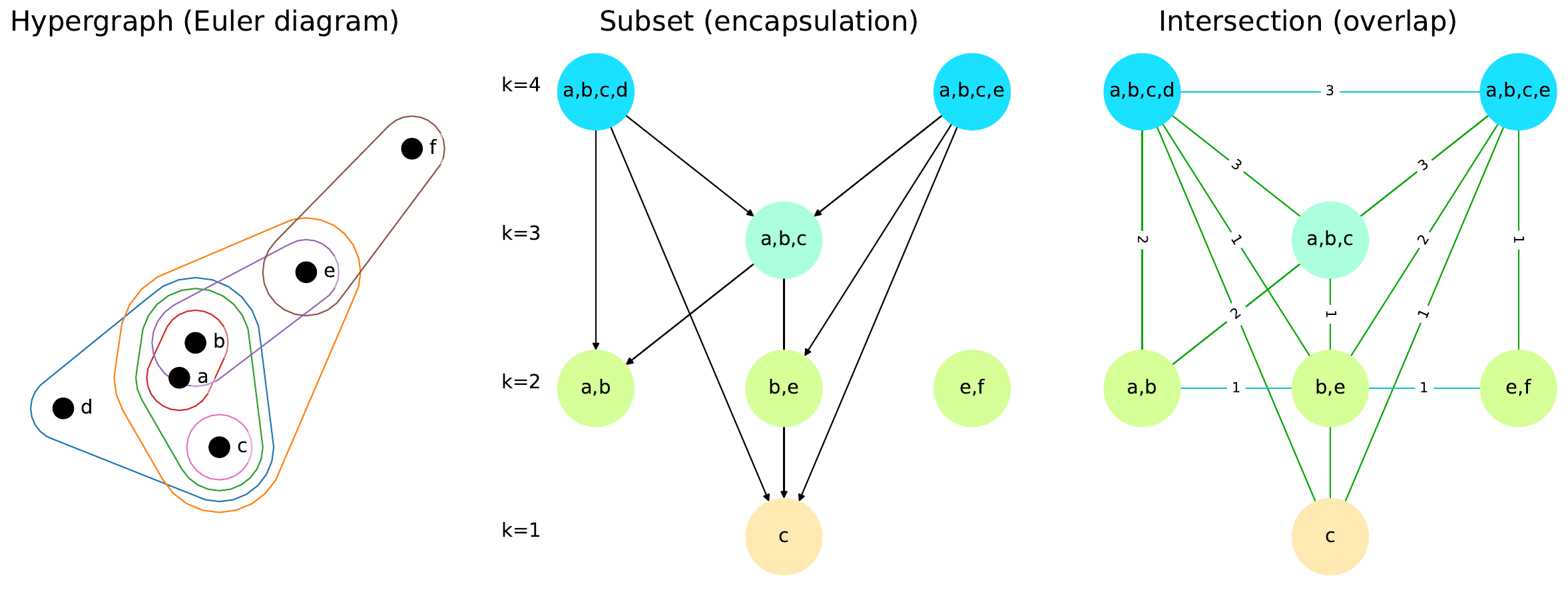}
    \caption{Examples of graphical representations of relations between hyperedges in a hypergraph, shown as an Euler Diagram in the left panel, drawn as sets of nodes in the center and right panel. We call the directed graph in the center an \emph{encapsulation DAG}, and the black edges represent subset relationships between hyperedges. We call the undirected graph on the right an \emph{overlap} or \emph{intersection} graph, with edges that correspond to intersection relationships between hyperedges. The edge weights in the overlap graph represent the size of the intersection between the hyperedges connected by the edge. The intersections that occur across hyperedge sizes, shown as green edges in the overlap graph, can be made directed (for consistency with the encapsulation DAG, larger hyperedges point to smaller) and correspond to an \emph{overlap DAG}.}
    \label{fig:overlap-example}
\end{figure}

Formally we  represent the multiway interactions as a hypergraph $H=(V, E)$ where $V=1, 2, ..., n$ is the set of $n$ nodes and $E=\{e_1, e_2, ..., e_m\}$ is the set of $m$ hyperedges representing interactions between the nodes in $V$, with the size of each interaction measured as the number of nodes and represented by $\ell_i = |e_i|$. 

To understand the extent of nestedness in the structure of a hypergraph, we build a line graph where the nodes are hyperedges and where there is a a directed link between two hyperedges if one is a subset of the other. These links represent what we call \emph{encapsulation} relationships between hyperedges. More formally, given two hyperedges $e_i$ and $e_j$ such that $\ell_i > \ell_j$, we say $e_j$ is encapsulated by $e_i$ if $e_j \subset e_i$. 

The line graph representing encapsulation relationships is a Directed Acyclic Graph (DAG) $D_{e_i,e_j}$ of $H$, where a directed edge from hyperedge $e_i$ to hyperedge $e_j$ means that $e_i$ encapsulates $e_j$. Since for every connected $e_i$ and $e_j$ we know that $\ell_{e_i} > \ell_{e_j}$, a cycle in this graph would imply that a smaller hyperedge encapsulates a larger hyperedge, which is impossible, thus the graph is always a DAG. We refer to this DAG as the \emph{encapsulation DAG} of a hypergraph. By construction, a hypergraph corresponding to a simplicial complex would have the maximum possible number of edges in the encapsulation DAG. The center panel of Figure~\ref{fig:overlap-example} shows an example of encapsulation DAG. The number of edges in the encapsulation DAG is the number of encapsulation relationships present in the hypergraph.  As we will show,  these structures are useful for studying dynamical processes where the spreading occurs at the hyperedge level.

The encapsulation DAG is closely related to a Hasse Diagram representing a partial ordering of a set of sets. However, a Hasse Diagram is transitively reduced by construction, meaning that an edge between two nodes is removed if there is an alternative path between the nodes. Hasse Diagrams with weights associated to their nodes have been used to define weighted simplicial complexes of hypergraphs, which were further used to predict the evolution and recurrence of small groups \cite{kim_weighted_2017, sharma_hypergraph_2020}. While we will examine transitively reduced encapsulation DAGs in Section~\ref{sec:dag-paths}, for consistency we will refer to the line graph with edges representing encapsulation relationships as an encapsulation DAG throughout.

The encapsulation DAG is just one way to build a line graph from a hypergraph. Other objects can be defined by considering other relations between hyperedges. 
An important  relation is the \emph{intersection} between the hyperedges, which defines an \emph{overlap graph}. Given two hyperedges $e_i$ and $e_j$, an undirected edge exists between them if $|e_i \cap e_j| > 0$, and the weight of the edge is the size of the overlap $|e_i \cap e_j|$ (or, alternatively, normalized as $\frac{|e_i \cap e_j|}{\min{\ell_i, \ell_j}}$). If we remove the edges between hyperedges of the same size and impose directionality on the remaining undirected edges, for example by directing edges from larger hyperedges to smaller, we obtain a DAG that we call an \emph{overlap DAG}. The right graph in Figure~\ref{fig:overlap-example} shows an example of the intersection relation. We note that overlap graphs are also related to clique-graph representations of pairwise networks \cite{evans_clique_2010}.

Let us make a short digression about dynamics here, a topic that we will cover in more detail in Section~\ref{sec:dynamics}.
The \emph{encapsulation} and \emph{intersection} relations capture different ways in which hyperedges may be related with each other, but they also have different implications for dynamical processes on the hypergraph. The \emph{intersection} graph is compatible with dynamics centered on the nodes of the hypergraph. One can think here, for instance, of a threshold model where all of the nodes in a hyperedge become activated if a certain number (or fraction) of its nodes are already activated. The intersection graph then provides us with information on how the activation of one edge may spread into others. 
Take the hyperedge $\{e,f\}$ in Figure~\ref{fig:overlap-example} for instance. In that case, activating the nodes in $\{e,f\}$ may result, depending on the details of the dynamical model, in activation of the hyperedges $\{a,b,c,e\}$ and $\{b,e\}$, and trigger a cascade of activations in the hypergraph. The picture is strikingly different in the encapsulation DAG where $\{e,f\}$ is disconnected and therefore has no impact on future activations. Indeed, from that perspective, it is not the fact that node $e$ is activated that matters, but instead that hyperedges encapsulated in others are activated. In other words, the encasulation DAG is more naturally associated to dynamics where the states are defined on the hyperedges, in a way reminiscent of the Hodge Laplacian for diffusive processes \cite{schaub2020random}. A more thorough discussion of the interpretation of this type of dynamics, and its simulation on both synthetic and real-world hypergraphs, will be given in Section~\ref{sec:dynamics}.

Computationally, we construct the encapsulation and overlap graph structures using the following algorithms. For both algorithms, we first assign each hyperedge a unique label and construct a mapping between each node and the hyperedges it participates in. We then loop over each hyperedge $\alpha \in E$, and for each node $u\in \alpha$ we add edges from $\alpha$ to other hyperedges $\beta \in E$ based on the relation we are interested in. In the intersection graph, this means adding edges from hyperedge $\alpha$ to other hyperedges  $\beta \in E, u\in \beta$ with the weight defined above. For the encapsulation DAG, we only add edges to hyperedges $\beta$ that are encapsulated by $\alpha$, meaning we add edges where $\beta \subset \alpha$. After repeating this loop for each node in $\alpha$, the out-neighbors of $\alpha$ represent all of the hyperedges in $E$ that have the relevant relationship with the $\alpha$.

The complexity of this construction has two terms. We first loop over all hyperedges $m=|E|$ to construct a mapping from hyperedges to labels, and a mapping from nodes to the hyperedges they are members of, which takes $O(m\cdot\ell_{\max})$ time, where $\ell_{\max} = \max_{e \in E}{\ell_e}$ is the maximum length of a hyperedge. Once the mapping is constructed, we again loop over all $m$ hyperedges to find encapsulation and overlap relationships. The worst case time for a loop is the size of the largest hyperedge $\ell_{\max}$ times the highest degree node $k_{\max} = \max_{u \in V} |\{e | u \in e; e\in E\}|$. This second term dominates the first and so the worst case running time is $O(m\cdot\ell_{\max}\cdot k_{\max})$.

\section{Encapsulation in Empirical Data}

In this section we introduce basic measurements of encapsulation relationships in some empirical hypergraph datasets, all of which were made available online with the publication of \cite{benson_simplicial_2018}. We focus in particular on coauthorship \cite{sinha_MAG_2015}, social contact \cite{stehl_contact_2011, mastrandrea_contact_2015}, and email communication datasets \cite{yin_local_2017, leskovec_evolution_2007}. In Table~\ref{table:datasets}, we show some statistics of the largest connected components of the hypergraphs. Following \cite{benson_simplicial_2018}, we exclude hyperedges of size greater than 25 nodes to keep some amount of consistency across the datasets. As mentioned above, we also ignore multiedges in the datasets and therefore consider the simple hypergraph representation of each.

\begin{table}[h]
\centering
\begin{tabular}{@{}lcccc@{}}
Dataset           &  $n_{cc}$ & $m_{cc}$ & Proj. Density & DAG Edges \\ \midrule
coauth-MAG-Geology     & 898648   & 947977   & $10^{-5}$ & 1650117      \\
coauth-MAG-History     & 219435   & 205531  & $10^{-5}$ & 217627      \\
contact-high-school       & 327      & 7818   & 0.11  & 7942         \\
contact-primary-school & 242      & 12704  & 0.29  & 16199        \\
email-Enron            & 143      & 1512    & 0.18 & 8240         \\
email-Eu               & 979      & 25008  & 0.06  & 277224       \\ \bottomrule
\end{tabular}
\caption{Number of nodes, number of hyperedges, the density in the projected graph, and number of DAG edges in the largest connected component of each empirical hypergraph after removing multiedges. The same statistics for the full datasets are available in \ref{app:data}.}
\label{table:datasets}
\end{table}

The coauthorship datasets, which include decades of published papers in multiple fields, contain numbers of nodes and edges that are multiple orders of magnitude larger than the face-to-face contact and email datasets. They are also orders of magnitude less dense in terms of the proportion of edges that exist in the projected graph where an edge exist between two nodes if they occur in the same hyperedge at least once.

\subsection{Degree in the Encapsulation DAG}
\begin{figure}
    \centering
    \includegraphics[scale=0.45]{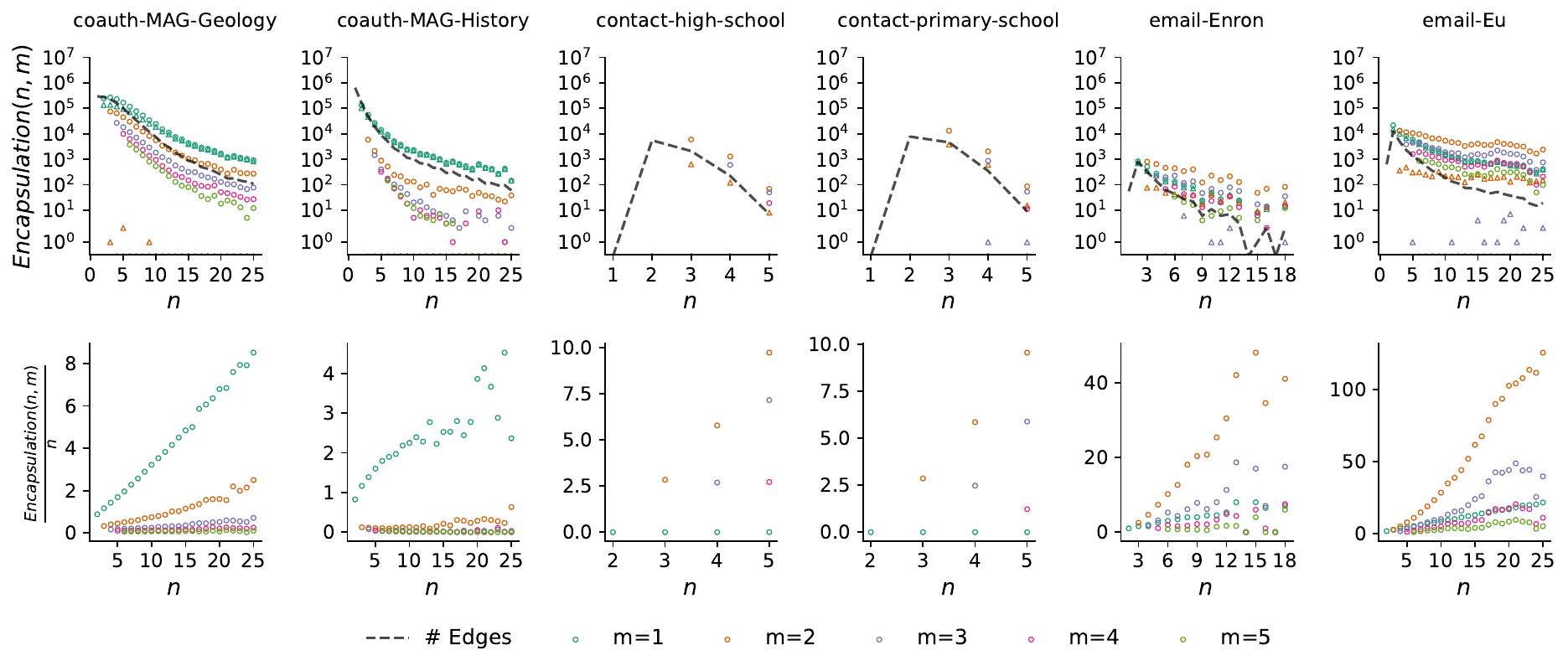}
    \caption{Encapsulation of size $m$ hyperedges by size $n$ hyperedges. The top row reports the number of size $m$ hyperedges that are encapsulated by size $n$ hyperedges as well as the number of size $n$ hyperedges for each $n$. Circles show the observed number of encapsulations, while triangles show the same counts for hypergraphs after applying the layer randomization described in the text. The bottom row shows the same quantity but normalized by the number of size $n$ hyperedges.}
    \label{fig:encapsulationmn}
\end{figure}

For each hyperedge, we are interested in the extent to which it encapsulates other hyperedges present in $E$, or equivalently we are interested in its out-degree in the encapsulation DAG. In the top row of Figure~\ref{fig:encapsulationmn}, we report the total number of hyperedges of each size $m$ that are encapsulated by hyperedges of larger sizes $n>m$. The total number of hyperedges of each size $n$ is shown as a dotted line. 

For each $m$, the number of observed hyperedges encapsulated decreases with $n$, but so does the number of size-$n$ hyperedges. To account for the distribution of hyperedge sizes, in the bottom row of Figure~\ref{fig:encapsulationmn} we report the same counts but divided by the number of size-$n$ hyperedges, giving us the number of encapsulated size-$m$ hyperedges per size-$n$ hyperedge.

\begin{figure}
    \centering
    \includegraphics[scale=0.21]{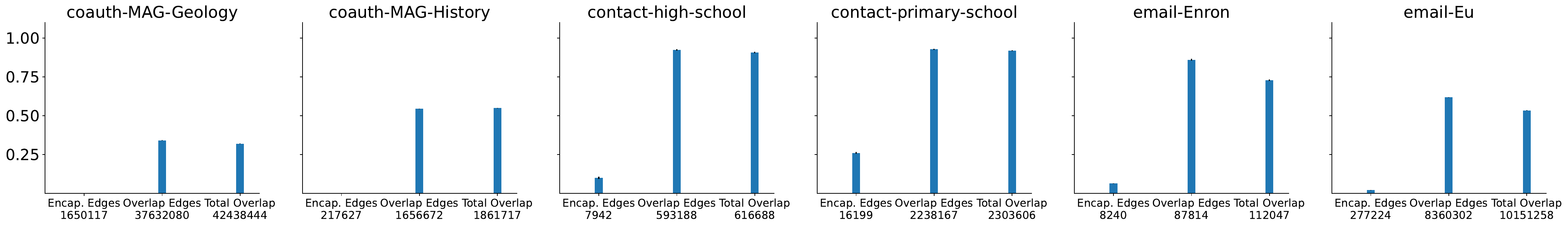}
    \caption{The effect of layer randomization on the number of edges in the encapsulation DAG, the overlap graph, and the sum of overlap graph weights. The vertical axis shows the average proportion of the relevant quantity in 5 layer randomization samples, with the observed quantities for each dataset reported in the axis labels. Across datasets, encapsulation relationships are substantially reduced, while overlap relationships are still maintained to a larger extent, especially in the face-to-face contact and email communication datasets.}
    \label{fig:layer-randomization}
\end{figure}

We also show the same quantity in a randomization of the hypergraph which we call the ``layer randomization''. The name comes from the fact that in this randomization procedure we view the sets of hyperedges of each size $k$ as a \emph{layer}, similar to the multiplex approach taken in \cite{sun_higher-order_2021}. The procedure then works as follows: for each layer of the hypergraph consisting of hyperedges of size $k$, we gather all of the hyperedges and the set of their constituent nodes, then shuffle the labels of the nodes. We repeat this procedure for every layer independently. The result is a hypergraph where the hyperedge size distribution and the unlabeled node degree distribution within each size layer are preserved, but the labeled node degree distributions within size layers, the node hyperdegree distribution, and, most importantly, the cross-size encapsulation and overlap relationships are randomized. In other words, we randomize the hypergraph across layers, but not inside layers. This is the reason why we opted for this randomization procedure, and not, for example, the configuration model for hypergraphs introduced by \cite{chodrow_configuration_2020}. Future work could investigate the effect of other randomization procedures such as those discussed in \cite{chodrow_configuration_2020, sun_higher-order_2021}. In Figure~\ref{fig:layer-randomization}, we show the proportion of encapsulation and overlap relationships destroyed by the layer randomization.

Only the coauthorship datests include hyperedges of size 1 ({\it{i.e.}}, nodes representing papers authored by a single individual). The number of encapsulations of 1-node hyperedges increases with $n$ after accounting for the number of size-$n$ hyperedges across all three datasets. This indicates that authors who are part of large collaborations also publish single author papers. However, the relationship is not as strong as would be expected under the simplex assumption. In that case,  every node should appear as a 0-simplex and the number of encapsulations would grow exactly as $y=n$ since all $n$ nodes would be encapsulated for every size-$n$ hyperedge. Instead the encapsulation relationship for single nodes grows sublinearly, indicating that there are many nodes which appear in hyperedges of size larger than 1, but never appear alone. Note also that the layer randomization does not substantially reduce the number of encapsulations of 1-node hyperedges, since the shuffling of the 1-node layer has no effect, and shuffling at each higher layer still results in some encapsulations of 1-node hyperedges necessarily, since the set of nodes in each layer does not change.

The relationship is even weaker for larger values of $m$. In this case, the simplex assumption would lead to the relationship $y = {n \choose m}$, since for every size-$n$ hyperedge all possible size-$m$ hyperedges would have to exist. However, for all values, the number of encapsulations per size-$n$ hyperedge stays well below 1, meaning that, on average, a size-$n$ hyperedge encapsulates few smaller hyperedges relative to its maximum capacity.
Notably, for all of the coauthorship datasets, encapsulation relationships tend to be destroyed  among hyperedges of any size after the layer randomization is applied, as expected.

The encapsulation structure of the face-to-face social contact hypergraphs appears to be more sparse than the rest of the datasets, partly due to the fact that there are fewer large interactions with a maximum interaction size of only 5 nodes. However, even with this more sparse structure, there are substantial encapsulation relationships, especially for hyperedges with 2 and 3 nodes.

The email communication hypergraphs show a substantially nested structure where large group emails are composed of groups with many smaller interactions in separate email chains, especially pairwise and 3-node interactions. This is consistent with an intuitive understanding of how email communication works within organisations: many small group email chains will naturally occur to facilitate day-to-day operations and side conversations, while large group emails will occur around big meetings, decisions, or announcements that involve larger proportions of the organisational structure. Interestingly, compared to the coauthorship data, the layer randomization keeps substantially more of the encapsulation relationships in the email communications. We hypothesize that this is due to the smaller number of nodes in the email datasets, which constrains the possible randomizations.

\begin{figure}
    \centering
    \includegraphics[scale=0.45]{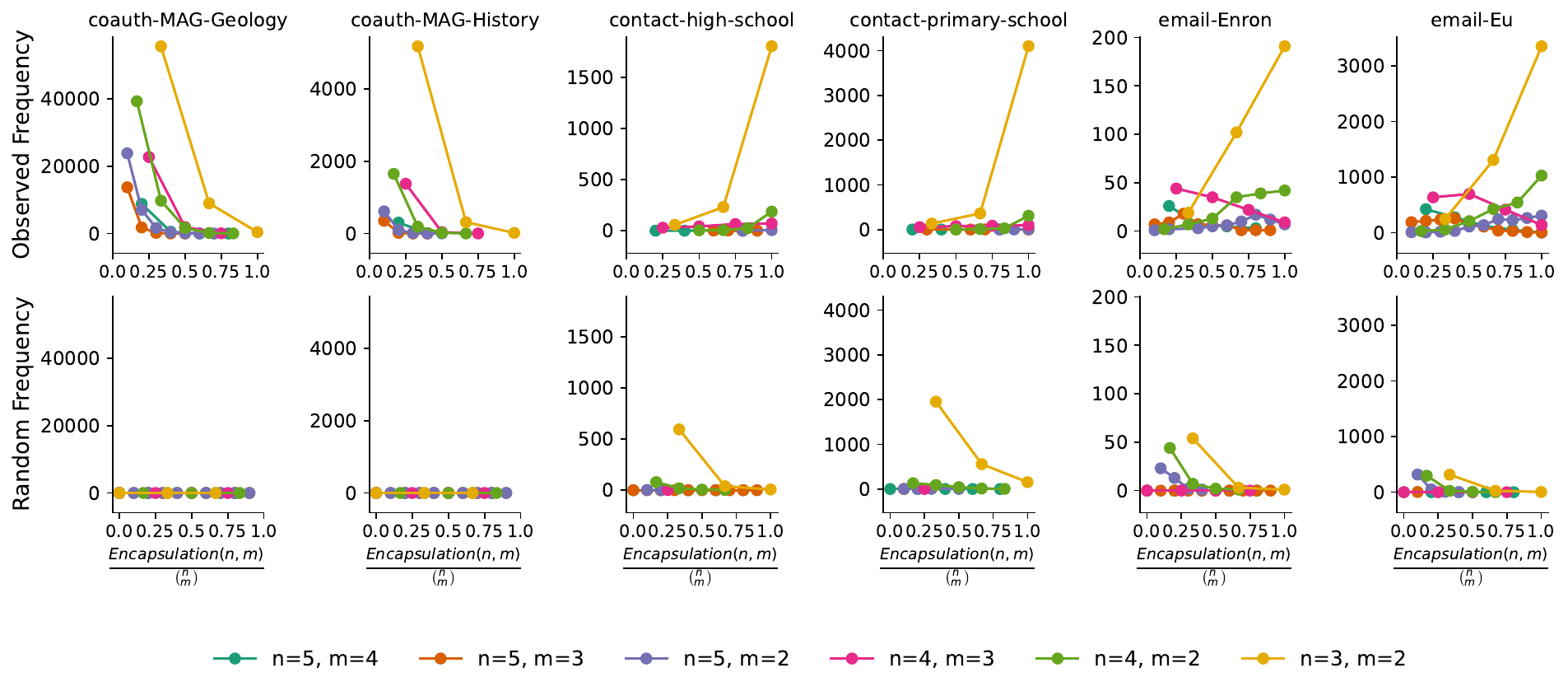}
    \caption{The distribution of number of encapsulations for each $(m,n)$ pair, where the horizontal axis is normalized by the maximum number of encapsulations per edge, $n \choose m$. If the simplex assumption were to hold empirically, each histogram should have one mode at 1.0, meaning that all size-$m$ hyperedges exist for every size-$n$ hyperedge. This is clearly not the case for the coauthorship datasets, where this pattern only holds for a relatively small number of ``closed'' triangles ($n=3, m=2$). In contrast, closed triangles are common in the fact-to-face contact and email communication hypergraphs, where the $n=3, m=2$ distribution is concentrated over 1.0. The bottom plots shows the same distributions in data randomized with the layer randomization. The randomizations have substantially fewer encapsulation relationships, but the effect of randomization is stronger in the larger, sparser coauthorship datasets than in the smaller and more dense social and communication networks.}
    \label{fig:overlapdists}
\end{figure}

In Figure~\ref{fig:overlapdists}, we show the distribution of encapsulation for each $(n,m)$ pair; that is, one distribution for each point in Figure~\ref{fig:encapsulationmn} up to $n=5$. We compute for each $\alpha \in E$ with $\ell_\alpha = n$ the number of out-neighbors of $\alpha$ in the encapsulation DAG that are of size $m$. We then normalize this quantity by the maximum number of subsets of size $m$, which is $n \choose m$. Thus if a histogram is fully concentrated on 1, there is full encapsulation and the simplex assumption holds. The bottom row of Figure~\ref{fig:overlapdists} shows the same histograms computed on the layer randomized version of the hypergraph.
As we observed in Figure~\ref{fig:encapsulationmn}, the number of encapsulations decreases for all of the coauthorship datasets when $n$ increases. The distributions in Figure~\ref{fig:overlapdists} show that the most common amount of encapsulation is exactly one subset (leftmost point of each line), and relatively few hyperedges fully encapsulate all of the possible subsets (rightmost point in each line). However, we observe the opposite pattern in the social contact and email datasets, where full encapsulation of 2-node hyperedges by 3-node and 4-node hyperedges is common in the observed data, and these relationships are destroyed by the layer randomization.

\subsection{Paths Through Encapsulation DAGs}
\label{sec:dag-paths}

In this section we show how analysis of encapsulation DAGs can help understand the structure of encapsulation relationships. An encapsulation DAG encodes interaction structure in at least 3 ways. As shown above, we can use the out-degree of a hyperedge in the DAG to measure the extent to which subsets of that hyperedge also appear as hyperedges. Similarly, the in-degree of a hyperedge in the DAG indicates the extent to which the supersets of a hyperedge exist, e.g., how much a given hyperedge is encapsulated. Finally, and this is the purpose of this section, the length of paths in the DAG indicates the ``depth" of encapsulation relationships. 

Here we analyze the height of \textit{rooted paths in the transitively reduced DAG}, inspired by the approach taken in \cite{vasiliauskaite_cycle_2022}. A rooted path is one that begins from a root node, which we define as a node in the DAG with zero in-degree and non-zero out-degree. We consider paths starting from root nodes because they indicate the maximum possible path lengths through the DAG. A transitively reduced DAG is one in which all edges representing shorter redundant paths are removed. For example, if we have the edges A-B, B-C, and A-C, in the transitively reduced DAG the edge A-C would be removed, since there would still be a path from A to C without that edge. Analyzing the DAG after removing these ``shortcut'' edges gives us a sense for the extent to which intermediate sized hyperedges are or are not present.

The distribution of path lengths in the transitively reduced DAG indicates the depth of the encapsulation relationships in the hypergraph. If the distribution is skewed towards the maximum length ($k-1$ edges for a hyperedge on $k$ nodes), this indicates a  hierarchy of encapsulations in the sense that  multiple intermediate hyperedges of different sizes are all encapsulated by the same larger hyperedge (the root). In contrast, if most path lengths are short, this indicates that encapsulation relationships in the hypergraph are concentrated between only two different sizes at a time, a kind of shallow encapsulation. Note that transitively reduced DAGs corresponding to two hypergraphs with very different encapsulation structures could have similar numbers of edges, but very different path length distributions. As we will discuss below, deeper and more hierarchical encapsulation relationships can have important implications for how a contagion can spread over the hyperedges of a hypergraph.

\begin{figure}
    \centering
    \includegraphics[scale=0.225]{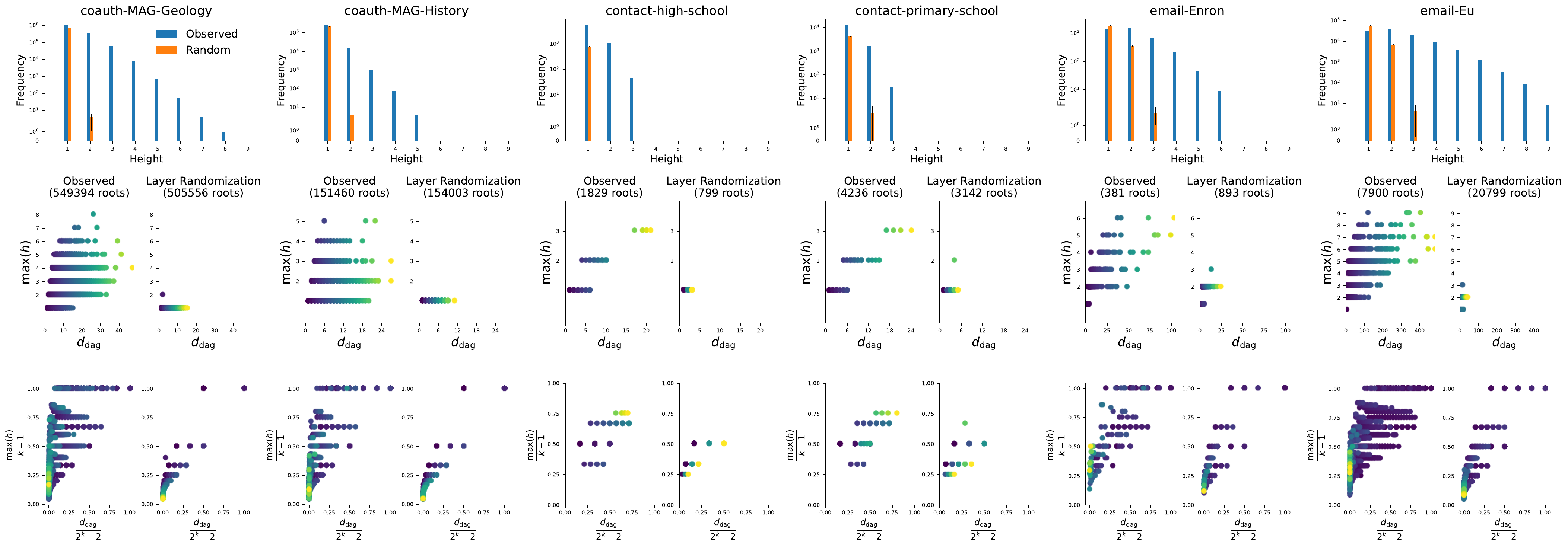}%
    \caption{Top: Distribution of heights of shortest paths in the encapsulation DAG starting from ``root" nodes, defined as DAG nodes (hyperedges) with zero in-degree and non-zero out-degree. Across datasets, the majority of paths are a single edge, while some are as long as 8 edges. After applying the layer randomization, the maximum path length drops to at most 3 edges across the datasets. Middle: Comparison of dag degree (horizontal axis) against maximum height path in the DAG (vertical axis) for all root nodes in observed DAG (left) and randomization (right). Colors represent absolute degree of the hyperedge in the DAG for comparison with the same quantities in the bottom row, normalized by their maximums. As DAG degree increases, longer paths are more likely. However, for the coauthorship and email datasets the majority of hyperedges with heights near the maximum are small (dark colors), consistent with the fact that the maximum path length is $k-1$. In contrast, in the social contact hypergraphs the hyperedges with high normalized DAG degree also have high absolute DAG degree and maximum path lengths.}
    \label{fig:dag-heights}
\end{figure}

In the top row of Figure~\ref{fig:dag-heights}, we show the distribution of heights in each dataset compared to the average over multiple layer randomizations. After randomization, the maximum path length through the transitively reduced DAG drops substantially in every dataset, and the number of paths of length 2 drops by multiple orders of magnitude in all of the coauthorship and contact datasets, but not in the email datasets.

In the middle and bottom rows of Figure~\ref{fig:dag-heights}, we plot for each root hyperedge its degree in the DAG against its maximum height (path length) in the transitively reduced DAG. The middle row shows the relationship without normalization for the observed (left) and layer randomized (right) hypergraph. The DAG degree of a hyperedge and its maximum length path in the transitively reduced DAG are positively correlated to varying extents across all of the datasets, but in the coauthorship datasets there are many hyperedges with high DAG degree that have relatively low maximum path lengths of only 2 or 3 edges.

As mentioned previously, the maximum height is bounded by $k-1$, where $k$ is the size of the hyperedge, since the maximum path length will pass through exactly one node (hyperedge) of each size $0 < k' < k$, of which there are $k-1$.  The bottom row of Figure~\ref{fig:dag-heights} again shows the relationship between DAG degree and maximum height, but with both quantities normalized by their maximums. 

As expected, when a root hyperedge has maximum degree in the DAG, it also has maximum path length (the opposite need not hold). The dark colored points in the top right of each normalized scatter plot indicate that only the hyperedges with small degrees have the maximum degree, meaning that they are also small hyperedges.

\subsection{Random Nested Hypergraph Model}

In this section we describe the Random Nested Hypergraph Model (RNHM) developed in \cite{kim_contagion_2023}, which we will use as a starting point for analyzing the relationship between nested hypergraph structure and a hyperedge contagion process. The parameters of the model are: the number of nodes $N$; the maximum sized hyperedge $s_m$; the number of hyperedges of size $s_m$, denoted $H_{s_m}$; and $\epsilon_s$, the probability of rewiring a hyperedge of size $s<s_m$.
Hypergraphs generated by this model are sampled by the following process. First, $H_{s_m}$ hyperedges of the maximum size $s_m$ are sampled, where the probability of a node being included in a hyperedge is uniform. Second, all of the subsets of those hyperedges (\emph{i.e.}, the powerset of every edge excluding sets with size less than 2) are added to the hypergraph. In some simulations, we also include all of the individual nodes as 1-node hyperedges. Finally, each of the encapsulated hyperedges with size $1<s<s_m$ are rewired with probability $1-\epsilon_s$, meaning that when $\epsilon_s$ is small, hyperedges of size $s$ are more likely to be rewired.

Rewiring a hyperedge involves (i) choosing a pivot node in the edge uniformly at random; (ii) deleting all other nodes from the edge; and (iii) replacing the deleted nodes with nodes chosen uniformly at random from outside of the hyperedges that are supersets of the original edge, ensuring that the new edge does not already exist in the hypergraph. Since this model will be used as a substrate for contagion dynamics in the next section, we further constrain the RNHM by rejecting hypergraphs that are not connected.

\begin{figure}
    \centering
    \includegraphics[scale=0.5]{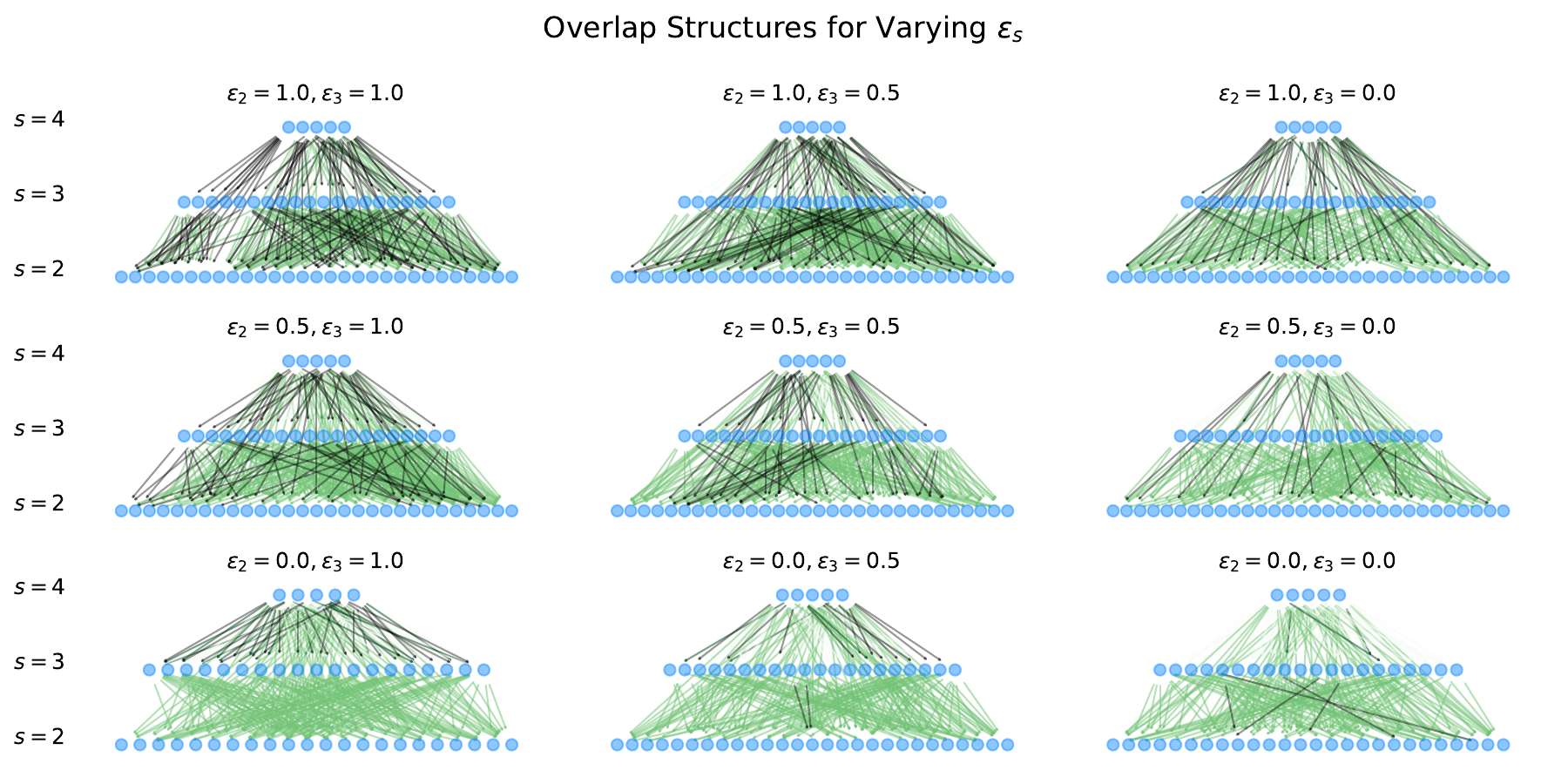}
    \caption{Example overlap structures for random nested hypergraphs with $N=20, s_m=4, H_{s_m}=5$ and varying the rewiring probabilities f$\epsilon_2$ and $\epsilon_3$. Black edges represent encapsulation relationships and green edges represent overlap relationships between hyperedges of different sizes. Note that every encapsulation edge is also an overlap edge.}
    \label{fig:nested-example}
\end{figure}

In Figure~\ref{fig:nested-example} we show DAG representations of random nested hypergraphs, where edges of the encapsulation DAG are drawn in black and edges from the overlap DAG are drawn in green. As $\epsilon_s$ decreases, so does the number of encapsulation relationships (DAG edges). When $\epsilon_s=1$, no hyperedges are rewired, so all encapsulation relationships exist. As $\epsilon_s$ decreases, rewiring of hyperedges reduces the number of encapsulation relationships until, when $\epsilon_s=0$, almost no encapsulation relationships between 4-node and $s$-node hyperedges exist. However, since the $s$-node hyperedges were constructed based on the set of nodes that appeared in the 4-node hyperedges, some encapsulation relationships may randomly remain after rewiring.

\section{The Role of Encapsulation Structure in Dynamics}
\label{sec:dynamics}

In this section we show that encapsulation plays a role in modulating the relationship between higher-order interactions and dynamical processes. We study a complex contagion process for which encapsulation and overlapping structures are vital to spreading. Our work builds on advances in the study of dynamical processes on higher-order structures, including the relationship between spreading dynamics on hypergraphs compared with simplicial complexes, where encapsulation relationships are implied \cite{petri_simplicial_2018, burgio_evolution_2020, landry_effect_2020, sun_higher-order_2021, zhang_higher-order_2023, st-onge_heterogeneous_2023, ferraz_de_arruda_multistability_2023}. It is important to emphasise that our analysis focuses on a purely higher-order effect, as the notion of encapsulation has no counterpart in classical networks.

We study a hypergraph complex contagion process where in each discrete timestep, every node $u\in V$ and hyperedge $\alpha \in E$ in the hypergraph is in a binary state, either inactive or active. We represent these states using two binary vectors, $s_u$ for nodes and $x_{\alpha}$ for edges, which both take a value of $0$ if the corresponding node or edge is inactive, and $1$ if active. At each time step, an inactive hyperedge $\alpha, x_{\alpha}=0$ is activated if more than a threshold $\tau$ of hyperedges which it directly encapsulates, {\it i.e.}, hyperedges of size $|\alpha|-1$, are also active. Therefore activation can only spread in hypergraphs with an encapsulation structure that is tightly nested, with many encapsulation relationships between adjacent layers of the DAG. We refer to this class of contagion as \emph{encapsulation dynamics} and focus on two variants depending on the influence we allow individual nodes to have on the dynamics.\footnote{Inspired by the language of topology, we may also call these dynamics \emph{subface} dynamics, referring to the fact that a subface of a simplicial complex would need to be activated for a larger face to activate.}

In the first variant, which we refer to as \emph{strict} encapsulation dynamics, individual nodes can only have influence in the dynamics if they appear in the hypergraph as a 1-node hyperedge. These 1-node hyperedges only appear in the coauthorship datasets, meaning that in the other datasets, individual nodes have no influence on the spreading process and their being in an active or inactive state has no bearing on the process beyond their participation in an active hyperedge that is encapsulated. In contrast, in the \emph{non-strict} variant we allow any individual node to influence pairwise interactions in which it participates. This corresponds to an assumption that all individual nodes are also 1-node hyperedges in the hypergraph and makes the state of individual nodes relevant to how the process can evolve. It also allows for exactly one kind of ``backwards'' activation, since activation of a large hyperedge will activate the individual nodes, while in general we do not allow activation of a large hyperedge to activate any of its subhyperedges in encapsulation dynamics. Instead, all activation flows upward through the encapsulation DAG from smaller to larger hyperedges. For a further discussion of the possible variants of encapsulation dynamics, see \ref{app:dynamics-discussion}.

Intuitively, encapsulation relationships are necessary to the spreading process in encapsulation dynamics, since larger hyperedges can only be activated if they encapsulate smaller hyperedges, which in turn must encapsulate still smaller hyperedges. We make an analogy between this process and building a campfire, where the smallest hyperedges correspond to dry leaves and twigs, medium hyperedges correspond to kindling, and the largest hyperedges correspond to the logs. Thus the ``goal'' of the encapsulation dynamical process we have defined is to catch the logs on fire by first lighting the fuel.

\begin{figure}
    \centering
    \includegraphics[scale=0.34]{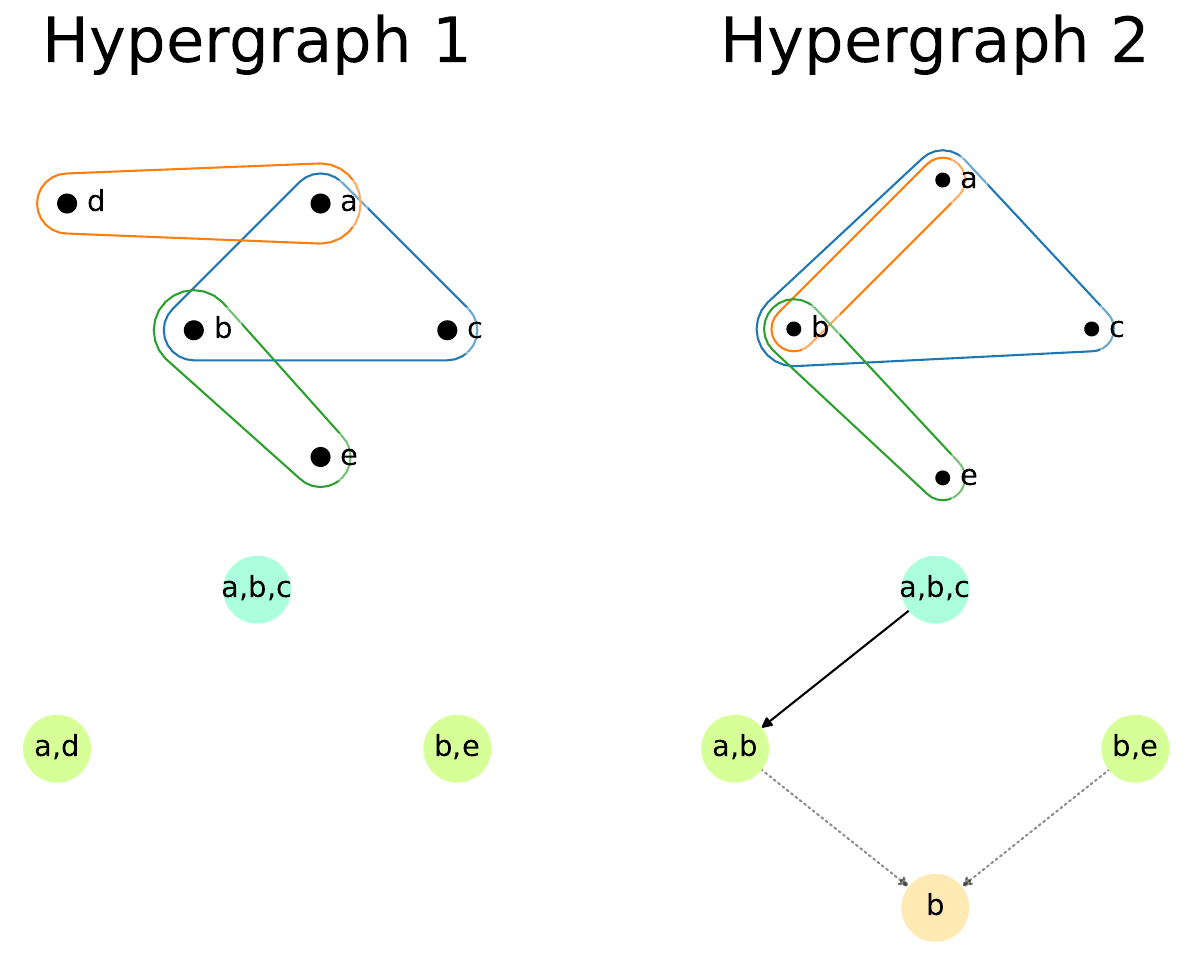}
    \caption{Two example hypergraphs that illustrate the difference between traditional node-based threshold dynamics and encapsulation dynamics. In Hypergraph 1, no matter where we place a seed-activated hyperedge, no other hyperedge can be activated in encapsulation dynamics because there are no encapsulation relationships. This is in contrast to node-based threshold dynamics where, for example, the seed activation of nodes $a$ and $b$ could lead to activation of all three hyperedges. In Hypergraph 2, there is an encapsulation relationship between hyperedges $\{a,b,c\}$ and $\{a,b\}$, meaning that as long as $\{a,b\}$ becomes active, $\{a,b,c\}$ can activate as well. In non-strict encapsulation dynamics, activation of node $b$ could also lead to activation of hyperedge $\{a,b\}$, even if $b$ is not in the hypergraph as an explicit 1-node hyperedge.}
    \label{fig:dynamics-example}
\end{figure}

The encapsulation dynamics can be seen as a generalisation of threshold models, which have been studied systematically in the context of opinion dynamics on graphs \cite{watts2002simple} and hypergraphs \cite{iacopini2019simplicial,de2020social,higham2021epidemics,noonan2021dynamics}. An important difference is that only activated nodes that are all connected by an active hyperedge can activate a larger hyperedge. 
From an opinion dynamics perspective, for instance, this could be interpreted as follows: a set of nodes that is part of a larger set may change the collective behavior only if nodes in the smaller set form an interacting unit, which allows them to coordinate their action.
In the illustration of Figure~\ref{fig:dynamics-example}, for instance, if we assume nodes $a$ and $b$ are activated in both hypergraphs, then their impact on node $c$ would be identical in the case of threshold models. The encapsulation dynamics distinguishes the two configurations, and the activation of node $c$ via the hyperedge $\{a,b,c\}$ is only possible when nodes $a$ and $b$ can coordinate their action via the encapsulated hyperedge $\{a,b\}$. In the non-strict encapsulation dynamics setting, where individual nodes are assumed to exist and have encapsulation relationships only with 2-node hyperedges, activation of node $b$ would also activate $\{a,b\}$ and lead to the activation of $\{a,b,c\}$. Thus we can view the non-strict variant of the dynamics as falling between the strict dynamics and node-based threshold models, where the existence and structural patterns of 2-node hyperedges are key to determining whether the non-strict dynamics behave more like strict or node-based threshold dynamics.\footnote{We also report simulations using more traditional threshold contagion dynamics based on node activations in ~\ref{app:threshold}.}

We simulate encapsulation dynamics by constructing the encapsulation DAG, but only keeping edges between hyperedges at adjacent layers, {\it i.e.}, where the difference in size is 1. In our simulations, we first place a given number of seed-activated hyperedges using one of the strategies described below. We then count for each hyperedge how many of its encapsulated hyperedges are seeds and deterministically simulate the dynamics forward. After each iteration, for every hyperedge $\alpha$ with size $\ell_\alpha$ nodes we update the number of its encapsulated $\ell_\alpha-1$ hyperedges that became activated. In practice, it is more efficient to update these counts by maintaining a reverse adjacency list of the encapsulation DAG so that we need only loop over the newly activated hyperedges and update the counts for the inactive hyperedges that they are encapsulated by.

We consider 4 different strategies for choosing seed hyperedges:
\begin{itemize}
    \item \textbf{Uniform}: Choose hyperedges uniformly at random.
    \item \textbf{Size Biased}: Choose hyperedges with probability proportional to their size ({\it i.e.}, choose the largest hyperedges first).
    \item \textbf{Inverse Size Biased}: Choose hyperedges with probability proportional to their inverse size ({\it i.e.}, choose the smallest hyperedges first).
    \item \textbf{Smallest First}: Explicitly choose the smallest hyperedges first. Practically, arrange the hyperedges in a vector ordered by increasing size, with hyperedges of the same size in random order. Choose seed hyperedges starting from the beginning of this vector.
\end{itemize}

We expect that in a hypergraph with deep encapsulation relationships the smallest first seeding strategy will be the most effective for strict encapsulation dynamics, since the small hyperedges must be activated or the dynamics will never reach the entire structure. In contrast, in non-strict encpsulation dynamics it may be the case that activating the largest hyperedges first will activate the most nodes that will in turn activate many pairwise hyperedges, potentially leading to more activation overall.

\subsection{Simulations on the Random Nested Hypergraph Model}
\begin{figure}
    \centering
    \includegraphics[scale=0.43]{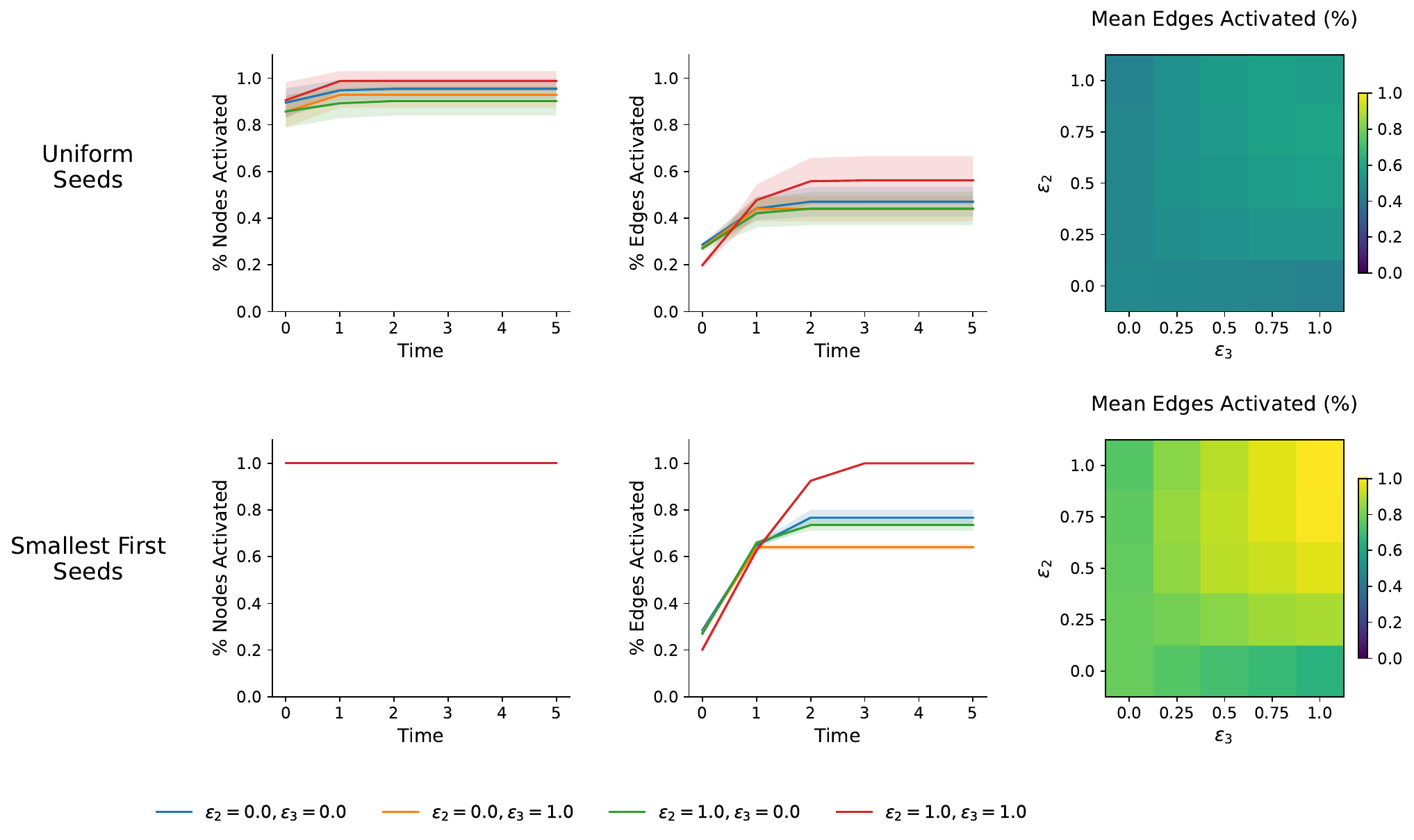}
    \caption{Comparison of encapsulation dynamics on random nested hypergraphs with varying combinations of $\epsilon_s$ for $N=20$ nodes, starting from  $H_{s_m}=5$ hyperedges of maximum size $s_m=4$ with $N$ seed hyperedges. The top row shows simulations with seed hyperedges chosen uniformly at random from all hyperedges, while the bottom row shows the smallest first strategy, which corresponds to choosing all of the individual nodes to activate. As expected, when all 1-node hyperedges are chosen as seeds, all hyperedges become activated when there is no rewiring since these hypergraphs have full encapsulation. The fewest edges are activated when $\epsilon_2=0$, meaning all hyperedges of size 2 nodes are rewired. The heatmaps in the right panel show the average proportion of edges activated for 16 parameter combinations, and shows that indeed $\epsilon_2$ and $\epsilon_3$ do not have symmetric impact.}
    \label{fig:RNHM-simulations}
\end{figure}

In Figure~\ref{fig:RNHM-simulations} we compare the encapsulation dynamics on random nested hypergraphs with varying combinations of $\epsilon_s$ for RNHM parameters $N=20, s_m=4, H_{s_m}=5$. In these simulations, we also include all of the individual nodes in the hypergraph. We show results using both uniform (top row) and smallest first (bottom row) seeding strategy, with number of seeds the same as the number of nodes $N$. Each point is an average over 50 realizations of the hypergraph and 50 simulations per realization. The smallest first strategy is more effective for all parameters, consistent with the ``campfire'' intuition of lighting the fuel to burn the logs.

In the smallest first simulations, all hyperedges are activated consistently when there is no rewiring of any hyperedges ($\epsilon_2=\epsilon_3=1$, red line), as expected. Interestingly, the dynamics are qualitatively different when either the 2- or 3-node hyperedges are rewired, but the other is left alone. More hyperedges are activated when only 3-node hyperedges are rewired ($\epsilon_2=1, \epsilon_3=0$, green line) compared to when only 2-node hyperedges are rewired ($\epsilon_2=0, \epsilon_3=1$, orange line). However, it is not the case that the most rewiring leads to the slowest activation dynamics. We attribute this to a combination between the stochasticity of the rewiring process and the relatively small number of nodes $N$, which can lead to situations where rewired hyperededges encapsulate each other randomly (see the encapsulation DAG in black for $\epsilon_2=0, \epsilon_3=0$ in Figure~\ref{fig:nested-example}, for example).

We also note that the smallest first seeding strategy as used in this setting would make node-based threshold dynamics trivial, since every node is activated in the seeding process. This illustrates the key conceptual difference between node-based and encapsulation-based dynamics: the latter requires explicit higher-order coordination among activated nodes, as well as encapsulation in the hypergraph structure.

\begin{figure}
    \centering
    \includegraphics[scale=0.45]{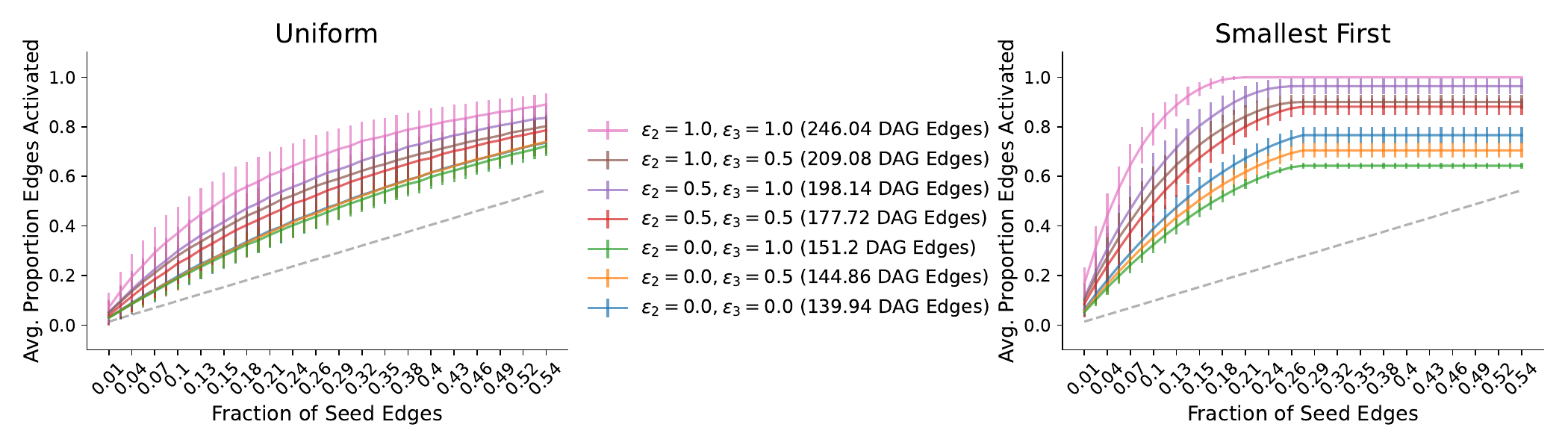}
    \caption{Simulating encapsulation dynamics on the RNHM shows that fewer seed edges are required when more encapsulation is present, and that the smallest first seeding strategy is more effective than uniform random. Using smallest first seeding, when there is no rewiring ($\epsilon_2=\epsilon_3=1$), meaning there is maximal encapsulation, the entire hypergraph becomes active consistently after about 20\% of hyperedges are activated as seeds. There is a clear separation between the outcomes depending on whether $\epsilon_2$ is non-zero.}
    \label{fig:RNHM-seeds}
\end{figure}

Figure~\ref{fig:RNHM-seeds} shows the average outcome of simulations on RNHMs with an increasing number of seed hyperedges again chosen with either uniform or smallest first strategy (25 realizations, 100 simulations per realization). In both cases there appear to be two distinct trends in the encapsulation dynamics results depending on whether $\epsilon_2$ is zero, meaning all 2-node hyperedges are rewired. Activation spreads to a larger number of hyperedges when $\epsilon_2 > 0$, consistent with the result from Figure~\ref{fig:RNHM-simulations}. When 2-node hyperedges are fully rewired, even with 50\% of edges being activated as seeds, only about 75\% of the total edges are activated by the end of the process in the best case.

\subsection{Simulations on Empirical Data}
\begin{figure}
    \centering
    \includegraphics[scale=0.37]{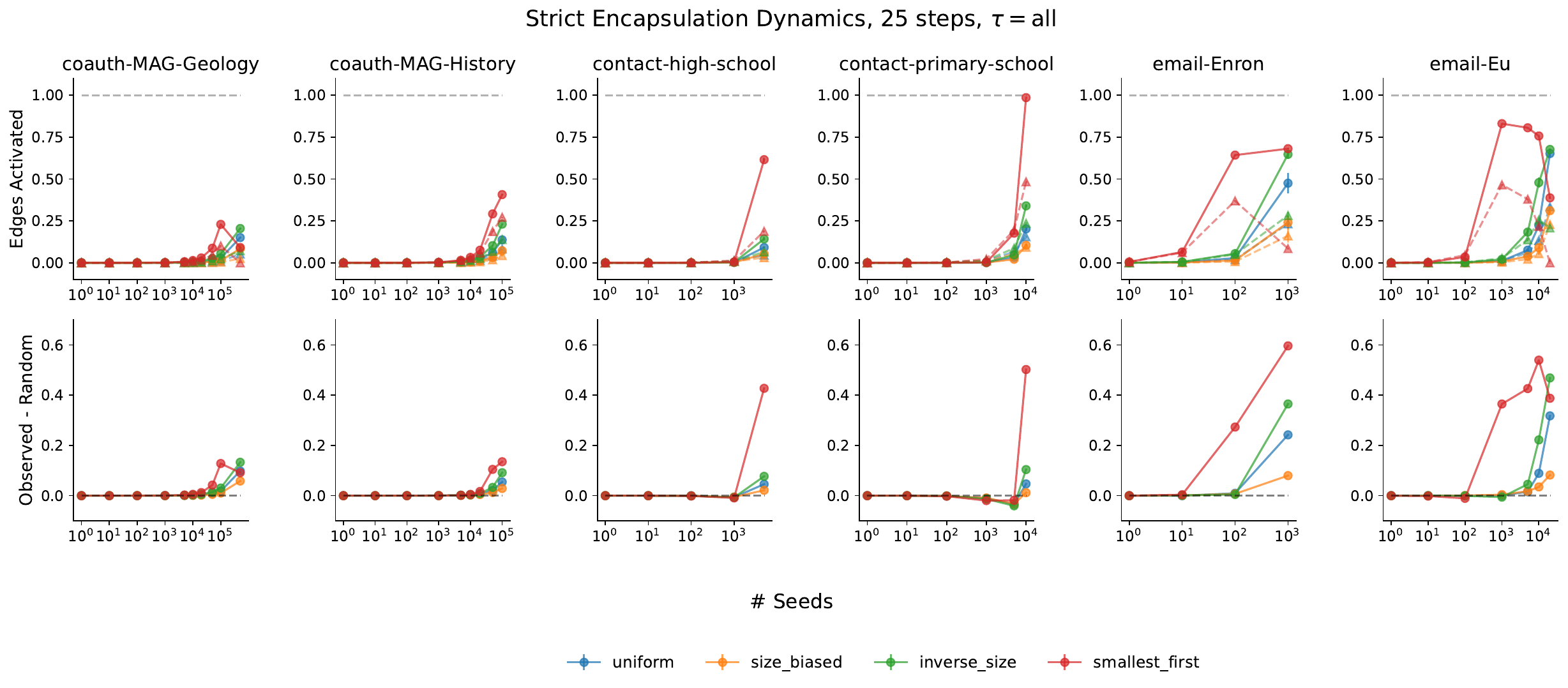}
    \caption{Simulation outcomes of strict encapsulation dynamics on empirical hypergraphs and their randomizations with varying seed strategies and number of seed hyperedges. The top row shows the average proportion of non-seed-activated hyperedges that are activated after 25 steps on observed hypergraphs (circles) and random hypergraphs (triangles). Error bars are too small relative to the axis scales to be visible. The bottom shows the difference between the proportions of activated hyperedges, observed minus random. In the strict setting, many seed activations are necessary for the dynamics to take off. Across datasets, the smallest first seed strategy is always most effective, followed by the inverse size strategy. This is in line with the ``campfire'' intuition behind the strict encapsulation dynamics: in order to light the large log hyperedges, the small bits of fuel must first catch fire. }
    \label{fig:empirical-seed-simulation-strict}
\end{figure}

\begin{figure}
    \centering
    \includegraphics[scale=0.37]{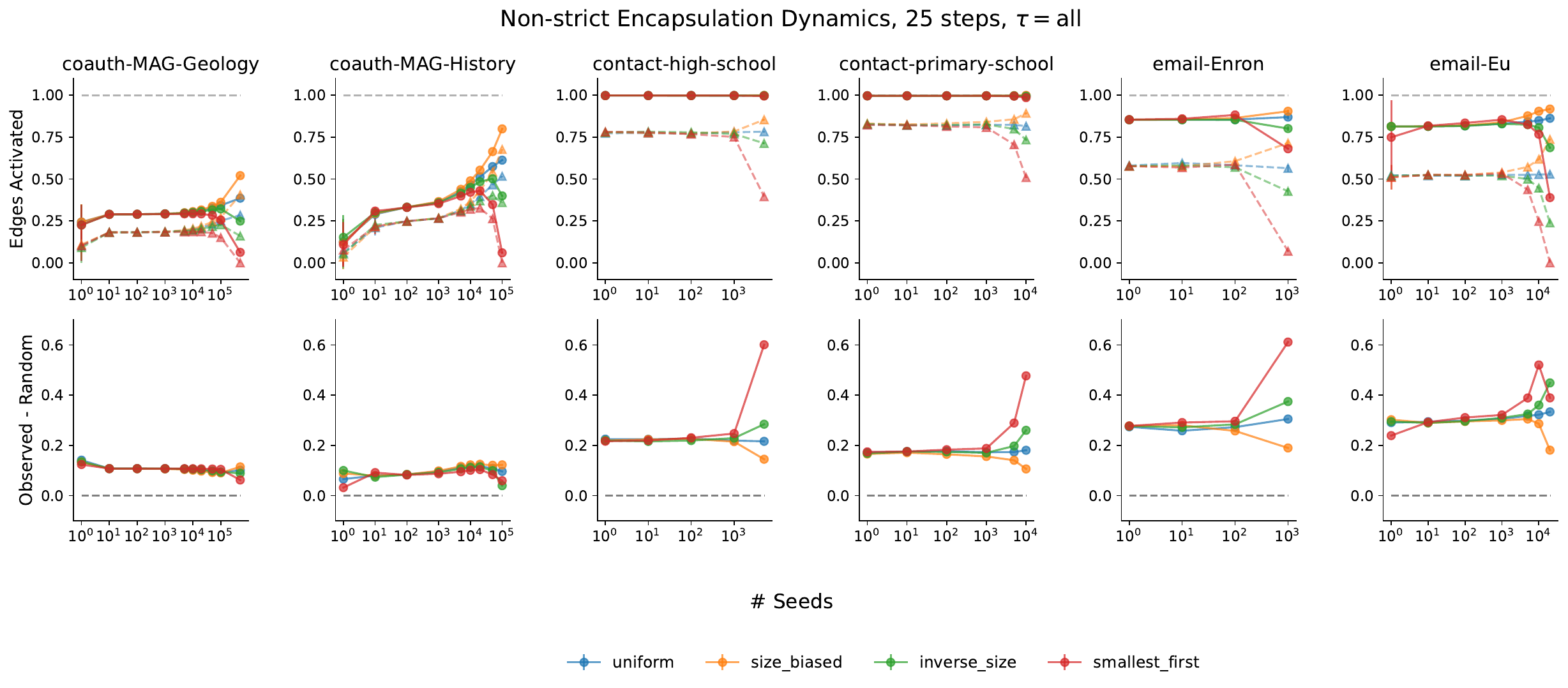}
    \caption{Simulation outcomes of non-strict encapsulation dynamics presented in the same format as Figure~\ref{fig:empirical-seed-simulation-nonstrict}. In the non-strict encapsulation dynamics, the seed strategies are similarly effective across datasets and number of seeds until very large numbers of seeds are chosen, in which case the most effective strategy varies across the datasets, except for the face-to-face social contact datasets where all seed strategies lead to full activation even with just 1 seed due to their substantial density. In the coauthorship datasets, choosing the largest hyperedges or choosing hyperedges uniformly at random are better than the strategies emphasizing smaller hyperedges at high numbers of seeds. While this may appear counterintuitive, in the non-strict setting activating a larger proportion of nodes can lead to many activations of pairwise edges in the first step, leading to more activation overall.}
    \label{fig:empirical-seed-simulation-nonstrict}
\end{figure}

We also simulated the encapsulation dynamics on the same empirical datasets described in Table~\ref{table:datasets} and their randomizations. In the top rows of Figures~\ref{fig:empirical-seed-simulation-strict} and \ref{fig:empirical-seed-simulation-nonstrict}, we show the proportion of non-seed hyperedges activated after 25 steps across all datasets with varying seed strategies and increasing number of initially active seed hyperedges.\footnote{Since the dynamics are deterministic once the seed hyperedges are chosen, usually only a small number of simulation steps are needed before the spreading stops. 25 steps is more than necessary for all of these datasets.} In the bottom rows of each figure, we show the difference between the observed and randomized outcomes.

In strict encapsulation dynamics (Figure~\ref{fig:empirical-seed-simulation-strict}), where pairwise edges can only be activated if one of their constituent nodes is present as a hyperedge, no further hyperedges are activated on average for small numbers of seeds across the coauthorship and face-to-face contact datasets. In the email datasets, the dynamics already take off with just 10 seed hyperedges and the smallest hyperedges first strategy clearly has an advantage in both the observed and randomized datasets. In fact, across all of the datasets the smallest first strategy is the most effective, and it also tends to be the strategy with largest difference in final activations between the observed and layer randomized hypergraphs. In general, activations on the layer randomization are much lower than in the observed hypergraphs, which is as expcted since the observed data contains many more encapsulation relationships.

In the non-strict encapsulation dynamics (Figure~\ref{fig:empirical-seed-simulation-nonstrict}), we again see that more non-seed edges are activated in the observed hypergraph with more encapsulation relationships. In the face-to-face social contact datasets, a single seed is enough to activate the entire observed hypergraph. Similarly, in the email datasets the final number of activations is consistent regardless of the number of seeds, until falling off at high numbers of seeds, likely due to the smaller proportion of available hyperedges to activate. However, in the layer-randomized hypergraphs, in the face-to-face contact and email datasets there appears to be a limit on the amount of non-seed hyperedges that can become activated.

We also note that in non-strict encapsulation dynamics, there is not a clearly best hyperedge seed placement strategy across the datasets. It is intuitive that the size biased strategies work well in non-strict dynamics with small numbers of seeds, since this strategy will by definition activate the most nodes, and these nodes can in turn activate pairwise edges they participate in, essentially translating into more seeds.

\section{Conclusion}

Higher-order networks have emerged, in recent years, as a promising approach to represent and model interacting systems. 
Among this broad family of models, approaches based on hypergraphs help to characterise the global structure and collective dynamics when interactions involve more than two agents. In this work, we have proposed novel ways to quantify the relations between hyperedges in real-world datasets. Based on the notions of overlap and encapsulation, we propose two alternative ways to represent a hypergraph as a graph where the nodes are the original hyperedges. In this {\em line graph} representation, edges may be directed to encode the encapsulation of a hyperedge in another, or undirected to encode the number of nodes in common between them. We have focused in detail on the structure induced by encapsulation, proposing a randomization strategy to erase encapsulation relations between hyperedges, while preserving other structural patterns, and quantifying how different real-world data are from what would be expected in a simplicial complex representation.

As a second step, we turned to dynamics. In contrast with  works focusing on the difference exhibited by a dynamical process on a hypergraph and on its corresponding projection on a graph, we explore the impact of encapsulation on spreading and compare the dynamics taking place on real-work hypergraphs and their randomization. To do so, we focus on a dynamical process specifically designed for hypergraphs -- the encapsulation dynamics is trivial on graphs -- and demonstrate that encapsulation facilitates spreading in situations when smaller hyperedges fuel the activation of larger hyperedges. Our work contributes to the recent efforts to understand how hypergraph structure impacts dynamics. Future research directions include a more thorough focus on the importance of overlap, but also testing our metrics to study other dynamical models, e.g. for synchronisation.

There remain many potential avenues for future work in this area. We have focused on a simple, size-layer-based approach to randomizing hypergraphs, but there exist in the literature other ways of randomizing hyperedges, including the configuration model approach introduced in \cite{chodrow_configuration_2020} and the multiplex approach in \cite{sun_higher-order_2021}. In contrast to our randomization, which preserves the size distribution of hyperedges and the unlabeled within-layer node degree distributions, both of these models preserve more general notions of degree, including the overall hyperdegree and the detailed within-layer degree of each node. Another potential research direction concerns the encapsulation dynamics, that was kept as simple as possible for the purpose of this work, but could be defined in different variants, as we allude to in \ref{app:dynamics-discussion}, in the same was that different types of threshold dynamical models have been explored in the literature. Finally, as we noted, the intersection and encapsulation relations are just two out of the several ways in which the relation between hyperedges can be measured. A combined analysis of the multiple line graphs that can be associated to the same hypergraph is also a promising research direction.

In this work, we ignored the temporal aspect of hypergraphs, however in the future the ideas introduced here could be extended to understand encapsulation patterns in temporal or dynamic hypergraphs, following work such as \cite{iacopini_temporal_2023}.
Our work could also be integrated with existing literature on higher-order motifs in hypergraphs \cite{lotito_higher-order_2022}. Further research could also be done on analyzing the DAG structures we investigated here using recent work on the cyclic analysis of DAGs \cite{vasiliauskaite_cycle_2022}.

\section*{Availability of Code and Data} Code implementing the measurements and simulations shown in this paper will be made available at \url{https://github.com/tlarock/encapsulation-dynamics/} \cite{larock_encapsulation_2023}. All of the empirical data was made available with the publication of \cite{benson_simplicial_2018} and can be found online at \url{https://www.cs.cornell.edu/~arb/data/}. 

\section*{Acknowledgment}
The authors acknowledge support from the EPSRC Grant  EP/V03474X/1. TL acknowledges the use of open source code made available by the developers of many projects including NumPy \cite{harris2020array}, SciPy \cite{2020SciPy-NMeth}, NetworkX \cite{SciPyProceedings_11}, MatPlotLib \cite{Hunter_2007}, and compleX Group Interactions (XGI) \cite{Landry_XGI_A_Python_2023}.

\appendix

\section{Data}
\label{app:data}
Table~\ref{app:table:datasets} shows the same statistics as Table~\ref{table:datasets}, but for the whole hypergraph, rather than just the largest connected component.

\begin{table}[h]
\centering
\begin{tabular}{@{}lcccc@{}}
\toprule
Dataset           & n       & m    & Proj. Density  & DAG Edges \\ \midrule
coauth-MAG-Geology     & 1256385 & 1203895 & $10^{-5}$ & 1666414  \\
coauth-MAG-History     & 1014734 & 895439  & $10^{-5}$ & 276588      \\
contact-high-school    & 327     & 7818  & 0.11  & 7942      \\
contact-primary-school & 242     & 12704 & 0.29  & 16199        \\
email-Enron            & 143     & 1512  & 0.18 & 8240         \\
email-Eu               & 998     & 25027 & 0.06  & 277224     \\ \bottomrule
\end{tabular}
\caption{Number of nodes, hyperedges, and DAG edges in each dataset after removing multiedges. In our measurements we include all hyperedges, while in simulations we focus on the largest connected components.}
\label{app:table:datasets}
\end{table}

\section{Discussion of Alternative Dynamics}
\label{app:dynamics-discussion}
Due to the multidimensionality inherent to hypergraphs, there are numerous valid choices for specifying a spreading process of the type we study here, each of which have their own conceptual and practical advantages and pitfalls. In this Appendix we discuss some of the potential alternatives that could be investigated in the future. We focus specifically on the specification of spreading over hyperedges - for a brief discussion of node-based threshold models on hypergraphs, see \ref{app:threshold}.

The first and most important choice in specifying the dynamics is deciding which hyperedges can influence one another. In the main text, we presented a model where only hyperedges at adjacent levels in the encapsulation DAG can influence each other, e.g. one in which only hyperedges of size $k-1$ can influence a hyperedge of size $k$. These are in some sense the most directly applicable to the ``ideal'' encapsulation DAG, since the dynamics directly spread over the DAG structure. However, we are also interested in how our spreading process unfolds on empirical hypergraphs, and we cannot know in advance whether the DAG connectivity will be suitable for spreading.

With this limitation in mind, we can also specify a version of encapsulation dynamics where we relax the condition from requiring immediately adjacent hyperedges to \emph{empirically adjacent} hyperedges, meaning that a hyperedge $\alpha$ can be influenced by hyperedges it encapsulates that are of the maximum size $k < |\alpha|$ existing in the hypergraph. For example, if a hyperedge on 4 nodes does not encapsulate any hyperedges on 3 nodes, but does encapsulate a hyperedge on 2 nodes, we allow this smaller hyperedge to influence the larger. 

The encapsulation dynamics presented in the main text are the most true to the spirit of the encapsulation relation, since they require that the encapsulation DAG has a specific structure. The empirical encapsulation relaxation is more flexible and compatible with the variety of structures we expect to see in empirical data, but the cost of this flexibility is that in some cases very small hyperedges can ``punch above their weight'' by activating much larger hyperedges just by virtue of being the only observed encapsulated edge.

We can address this issue in a few ways. In the first place, we could set the threshold $\tau$ to be at least the number of individual nodes in the hyperedge. With this threshold, it would only be possible for single nodes to activate a larger hyperedge if all of them were activated. However, this ``global'' threshold could have the effect of making it impossible to activate some hyperedges, for example a hyperedge with only one encapsulation, but where that encapsulation is of size $k-1$, which would also be counter-intuitive. Instead, size-specific threshold models could be given, such that a different number of different sized hyperedges could be necessary to activate a hyperedge.

Finally, there is the question of whether activation should go in only one direction, from smaller hyperedges up to larger hyperedges, or in both directions. In this work we have only allowed activation to flow from smaller to larger hyperedges, but it would be equally reasonable to assume that once a larger hyperedge has been activated, all or some of its subsets also become active. We leave investigation of this style of model for future work.

\section{Threshold Contagion Model}
\label{app:threshold}
\begin{figure}
    \centering
    \includegraphics[scale=0.3]{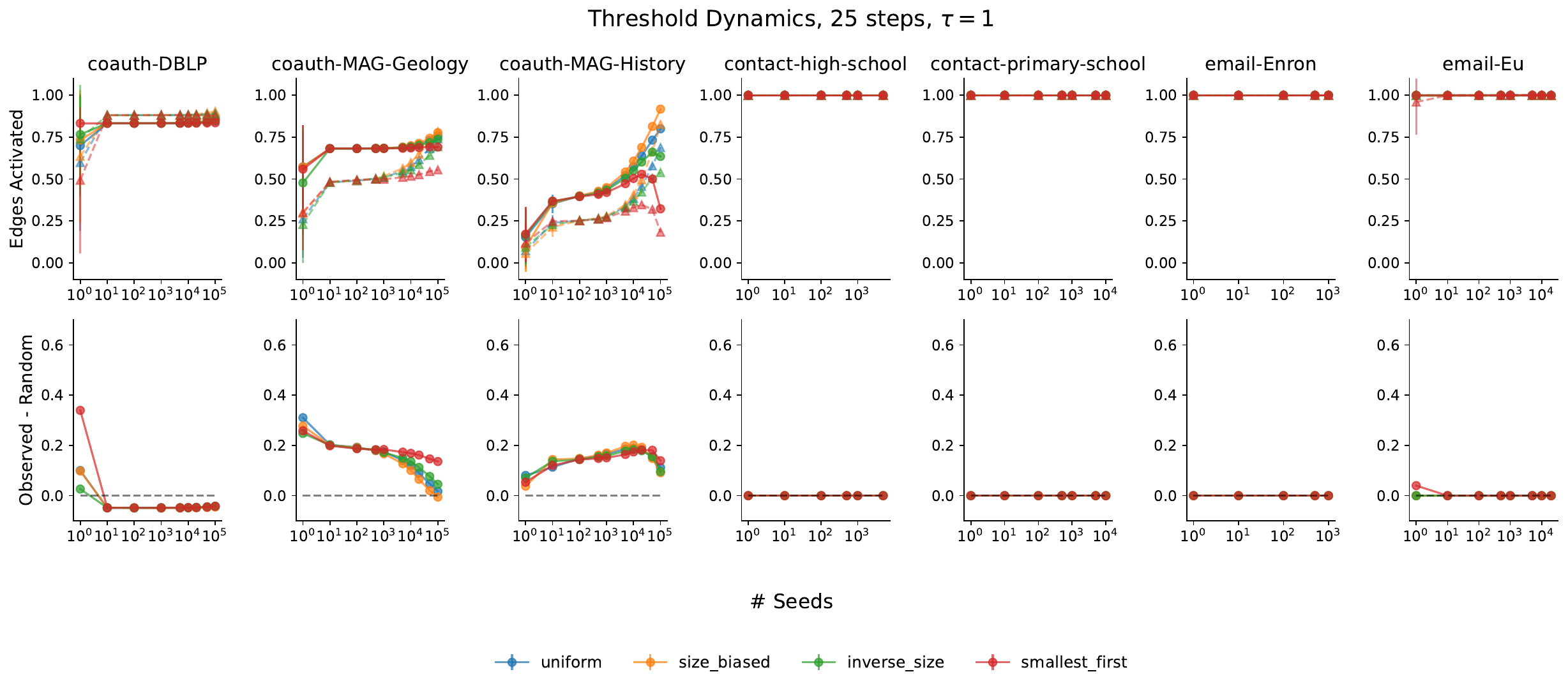}
    \includegraphics[scale=0.3]{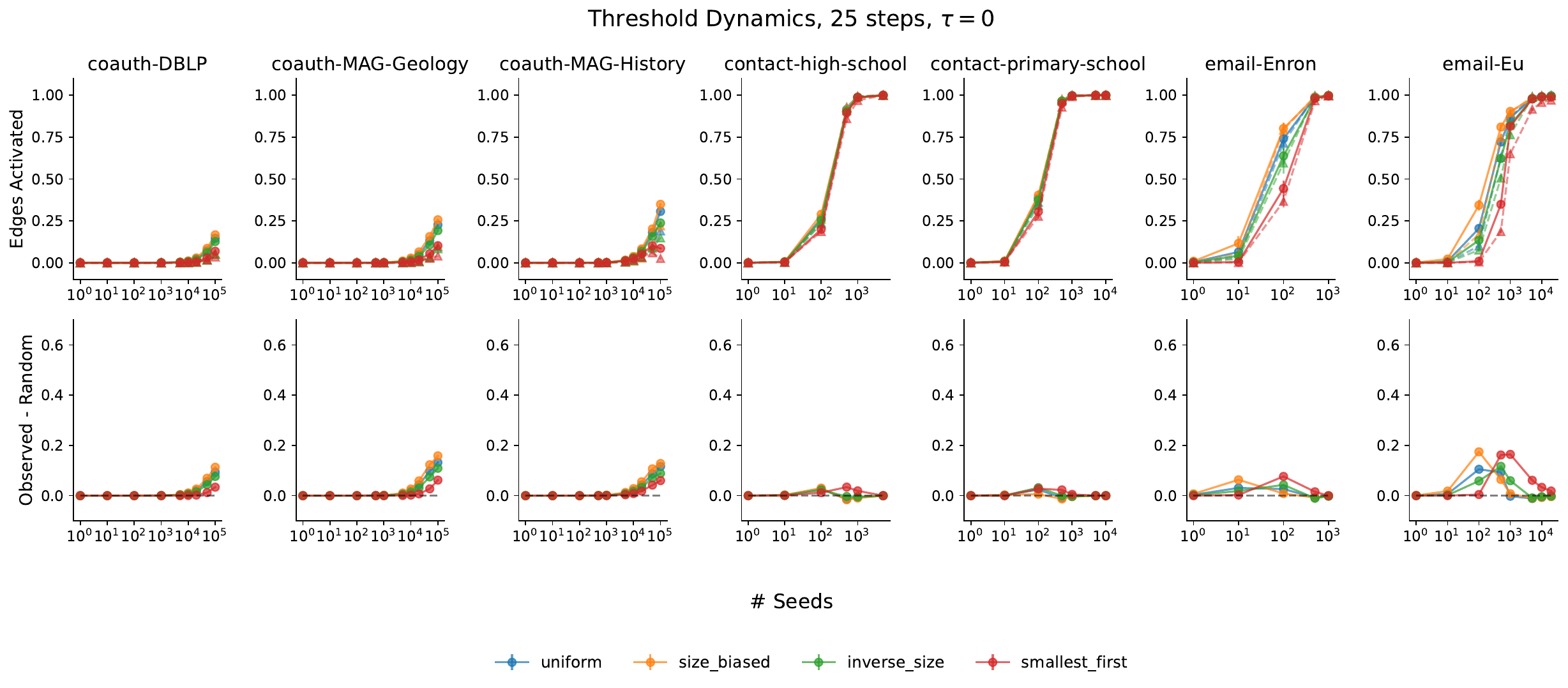}
    \caption{Proportion of non-seed hyperedge activations in node-based threshold dynamics with two different thresholds. When $\tau=1$, an inactive hyperedge becomes active if all but 1 nodes in the hyperedge are active. When $\tau=0$, an inactive hyperedge becomes active only when all of its constituent nodes have become active.}
    \label{app:fig:empirical-seed-simulation-down}
\end{figure}

In this Appendix, we show some results on a traditional node-based threshold contagion model on a hypergraph to contrast with the encapsulation dynamics we introduced in the main text. Just as in encapsulation dynamics, in our threshold model every node $u\in V$ and hyperedge $\alpha \in E$ in the hypergraph is in a binary state, either inactive or active, in each discrete timestep. At each step, an inactive hyperedge $\alpha, x_{\alpha}=0$ is activated if the number of already-activated nodes within the hyperedge is larger than a threshold. When a hyperedge is activated, all of its member nodes $u \in \alpha$ are also activated. We define the threshold based on the size of the hyperedge, specifically $|\alpha|-\tau$. An inactive hyperedge $\alpha$ will be activated if $$\sum_{u \in \alpha} s_u \geq |\alpha| - \tau,$$ that is, if the number of activated nodes is greater than the size of the hyperedge minus the threshold.

These dynamics could still be sensitive to encapsulation structure in a hypergraph, however the overlap structure of the hypergraph can play an equally important role, since there is no requirement that smaller hyperedges are activated first to activate enough nodes to finally activate larger hyperedges.

We run simulations on empirical datasets using two threshold values: $\tau=0$ and $\tau=1$ and present the results in Figure~\ref{app:fig:empirical-seed-simulation-down}. When $\tau=1$ (top plot), meaning that an inactive hyperedge $\alpha$ becomes active when the number of inactive nodes remaining in $\alpha$ is 1, a single seed activates the entire hypergraph for both the face-to-face contact and email datasets. In the coauthorship datasets, full activation is never achieved in either observed or randomized datasets.

When $\tau=0$, meaning all nodes must be activated for a hyperedge to become active, a sort of unanimity condition, the outcomes are dependent on the dataset. Starting with the email-Eu dataset, we see that as the number of seed hyperedges increases, choosing the largest hyperedges first is the most effective strategy on the observed data until the number of seeds increases passed $10^3$, where all of the methods converge. In the email-Enron dataset there is a similar pattern, but the difference between the outcome on the observed hypergraph and the random hypergraph is smaller across the simulations. The two contact datasets show similar patterns across all of the seeding strategies and in both observed and randomized hypergraphs, with full activation being achieved for the largest numbers of seeds. Finally, in the coauthorship datasets almost no activation occurs until more than $10^4$ hyperedges are activated as seeds, and choosing seeds proportional to their size is the best strategy.

\bibliographystyle{abbrv}
\bibliography{references}
\end{document}